\documentclass[prb,twocolumn,11pt,showpacs]{revtex4}
\usepackage{epsfig}
\usepackage{graphicx}

\begin{document}

\title{Finite-size effects in microrheology}
\author{I. Santamar\'{\i}a-Holek\dag, J. M. Rub\'{\i}\ddag}
\affiliation{\dag Facultad de Ciencias, Universidad Nacional
Aut\'{o}noma de M\'{e}xico.\\
Circuito exterior de Ciudad Universitaria. 04510, D. F., M\'{e}xico}
\affiliation{\ddag Facultat de F\'{\i}sica, Universitat de Barcelona.
\\
Av. Diagonal 647, 08028, Barcelona, Spain}

\begin{abstract}
We propose a model to explain finite-size effects in intracellular
microrheology observed in experiments. The constrained dynamics of the
particles in the intracellular medium, treated as a viscoelastic
medium, is
described by means of a diffusion equation in which interactions of the
particles with the cytoskeleton are modelled by a harmonic force. The
model
reproduces the observed power-law behavior of the mean-square
displacement
in which the exponent depends on the ratio between
particle-to-cytoskeleton-network sizes.
\end{abstract}

\maketitle

\section{INTRODUCTION}

Transport of particles as vesicles, lipid granules or chromosomes
through
the intracellular environment of eukaryotic cells has been subject of
intense theoretical and experimental research in recent years.\cite%
{blockNature,fabry,Nature98,peterman,lubensky,caspi,caspi3,mck1,wong,gittes,weitz82,cromosomas}
In these systems, the presence of molecular motors as well as of the
cytoskeleton may introduce strong spatial inhomogeneities and elastic
forces
which are responsible for anomalous transport of the particle.

A manifestation of this anomalous transport, observed through
video-based
and diffusing wave spectroscopy methods, is the power law behavior of
the
mean-square displacement (MSD) as a function of time.\cite%
{gittes,mck1,wong,caspi,caspi3,fabry,weitz82,cromosomas,marchesoniFinitesize}
These microrheology experiments have shown that the power-law exponent
depends
on the ratio between the size of the particles to the characteristic
length
of the polymer network constituting the cytoskeleton.\cite%
{fabry,gittes,wong,caspi} There are two limiting cases. The first one
in
which the radius of the particle $a$ is larger than the charateristic
length
of the polymer network $\xi$, $a/\xi > 1$; and the second one in which
the
radius of the particle is smaller than $\xi$, $a/\xi < 1$.

The results of video-based microrheology experiments in {\it in vitro}
F-actin networks shown in Fig. \textbf{1} are
an example of the MSD behavior in the second situation. It is important
to stress that
these results are consistent with those obtained from living cells and
\textit{in vitro} microtubule arrays by using also video-based methods,
in
which the MSD of probe particles also shown subdiffusion with different
exponents.\cite{caspi,caspi3}

\begin{figure}[tbh]
{}
\par
\centering \mbox{\resizebox*{7.50cm}{!}{\includegraphics{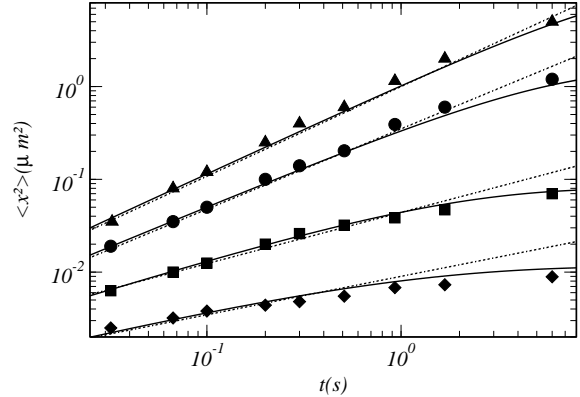}} }
\par
{\footnotesize {\ } \vspace{0.0cm} }
\caption{ MSD vrs time as obtained from the proposed model for
different particle-to-network-size ratio $a/\protect\xi$. Solid lines
correspond to the solution of Eq.(\protect\ref{xx-cromosomas}) with
(\protect
\ref{tau2}) whereas dotted lines represent the power-law given in Eq.
(%
\protect\ref{xx-3}) and valid at short times. Experimental data
(symbols) have been
taken from Ref. [8] and correspond to particles with radius $%
a=0.25 \protect\mu m$ and F-actin characteristic network sizes: $0.75
\protect\mu m$ (triangles), $0.55 \protect\mu m$ (circles), $0.30
\protect%
\mu m$ (squares) and $0.25 \protect\mu m$ (diamonds).}
\label{msd}
\end{figure}

In this article, we give an explanation of this effect in the case when
the
particles have a characteristic length smaller than the characteristic
length of the polymer network. We propose a diffusion model, taking
into
account the viscoelastic and bounded natures of the intracellular
medium, to
analyze the behavior of the MSD through a non-Markovian Brownian
dynamics
\cite{adelman,oxtoby,oliveira,oliveira2,nonmarkov,slow,polonica,SteadyState,balescu}
which considers
particle finite-size effects and confination.

At short times, our results lead to a power law behavior for the shear
modulus $G^{\prime \prime }(\omega)$ as a function of the frequency
$\omega$
in which the exponent is a function of the particle-to-network-size
ratio $%
a/\xi$. We also show the relation between the measured mean square
displacement
with an \textit{effective diffusion coefficient} $D(t)$ and an
\textit{effective friction} $\beta(t)$. It is necessary to stress that
these coefficients, containing
memory effects through its dependence on time,\cite{adelman,oxtoby} are
related with the position relaxation function of the Brownian particle
and,
only in the Markovian approximation, they reduce to the usual diffusion
and
friction coefficients.

The article is organized as follows. Section \textbf{II} will be
devoted to
introduce the generalized diffusion equation for a Brownian particle of
finite size and the main ingredients of the model. In Sec. \textbf{III}
we
apply the model to describe results of microrheological experiments in
the
intracellular media and \textit{in vitro} systems. In the discussion
section
we will summarize our main results.

\section{The generalized diffusion model}

For times much larger than the characteristic time $\beta
_{0}^{-1}=m/6\pi
a\eta _{s}$, where $a$ is the radius of the particle, $m$ its mass and
$\eta
_{s}$ the viscosity of the solvent, the dynamics of the Brownian
particle
through a viscoelastic medium can be described in terms of a
non-Markovian
diffusion equation of the form
\cite{slow,nonmarkov,SteadyState,balescu}
\begin{equation}
\frac{\partial \rho }{\partial t}=D(t)\nabla ^{2}\rho -\beta
^{-1}(t)\nabla
\cdot \left[ \rho \vec{F}(\vec{x})\right] ,  \label{Smol-Osc}
\end{equation}%
in which the effective diffusion coefficient $D(t)$ is related to the
effective mobility $\beta ^{-1}(t)$ through
\begin{equation}
D(t)=\frac{k_{B}T}{m}\beta ^{-1}(t),  \label{Stokes-Einstein}
\end{equation}%
where ${k_{B}T}$ is the thermal energy. In Eq. (\ref{Smol-Osc}), $\rho
(\vec{%
x},t)$ is the probability density, and $\vec{F}$ the force accounting
for
the interactions of the particle with the viscoelastic medium. It
contains
the elastic forces exerted by the polymer network over the particle and
takes
into account finite-size effects in the dynamics of the particle.

In the absence of molecular motors the dynamics of the particle is
subdiffusive.\cite{gittes,caspi,wong} In this case one may assume,
within
an effective medium approximation, that its motion is coupled to the
medium
through a harmonic force
\begin{equation}
\vec{F}(\vec{x})=-\omega _{0}^{2}\epsilon \vec{x},  \label{FB1}
\end{equation}%
where $\omega _{0}$ is a characteristic frequency of the polymer
network.

The parameter $\epsilon $ incorporates the mentioned finite-size
effects of
the particles. In particular, it reflects the effects of the stresses
exerted by the fluid over the
surface of the particle during its motion though the fluid. Its general
form,
$\vec{\vec{\epsilon}}\cong \frac{m}{6k_{B}T}a^{2}\vec{\vec{\beta}}\cdot
\vec{%
\vec{\beta}}$, has been calculated in Ref. [20] for the case
of a particle moving into a fluid under flow conditions by taking into
account the Fax\'{e}n theorem \cite{mazur-bedo,brenner}: the drag force
over
the particle is $\vec{F}_{drag}\propto
\vec{v}-\overline{\vec{v}_{0}}^{S}$
with $\vec{v}$ the velocity of the particle, and
$\overline{\vec{v_{0}}}%
^{S}\sim \vec{v}_{0}+\frac{a^{2}}{3!}\nabla ^{2}\vec{v}_{0}$ the
average of
the velocity of the fluid $\vec{v}_{0}$ over the surface of the
particle.\cite{SteadyState} For convenience, we will assume that $\epsilon
=\tau _{D}\beta (t_{0})$ with $\tau _{D}={ma^{2}\beta
(t_{0})}/{6k_{B}T}$
the characteristic diffusion time over the distance
$a$.\cite{SteadyState}
Therefore, the form of the elastic force (\ref{FB1}) then assumes that
it
originates from the fact that
particles have finite size and that point particles are not affected by
the
network in the time interval we are considering.

It is important to mention that the non-Markovian character of equation
(\ref%
{Smol-Osc}) is incorporated through the time dependence of the
effective
mobility $\beta ^{-1}(t)$. This time dependence can be shown to arise
from a
description given in terms of a generalized Langevin equations,\cite%
{adelman,oxtoby,oliveira,oliveira2,nonmarkov,slow,polonica} or from a
slow-varying-field
approximation of the corresponding equation obtained from the theory of
continuous time random walks.\cite{balescu} Also important to mention
is
the fact that, in the general case, the effective mobility
$\vec{\vec{\beta}}%
^{-1}(t)$ is not the usual transport coefficient of the Markovian
approximation. However, it can be shown that it recovers its usual form
in
the corresponding approximation.\cite{adelman,nonmarkov}

During the motion of the particle through the viscoelastic fluid,
hydrodynamic interactions with the polymer network introduce
corrections to
the mobility of the particle which are similar in form to those arising
form
the presence of other particles or from the presence of a wall or a
liquid-like membrane.\cite{brenner,saarloos,dufresne,bickel} For
example,
in the presence of a wall, these corrections depend on the ratio $a/h$
in
the form: $\beta _{wall}^{-1}\simeq \beta _{0}^{-1}\left(
1-0.625a/h\right) $
with $h$ the distance to the wall.\cite{brenner} It is then plausible
to assume
that the distance of the particle to the "wall" of the "cage" formed by
the
polymer network is of the same order as the characteristic length of
the
polymer network $\xi $, and then write in first approximation%
\begin{equation}
\beta ^{-1}(t)=\beta _{0}^{-1}\left( 1-B_{1}\frac{a}{\xi }\right)
\tilde{%
\beta}^{-1}(t),  \label{friccion-pared}
\end{equation}%
where $\tilde{\beta}^{-1}(t)$ is the dimensionless function of time
accounting for the memory
effects.\cite{adelman,oxtoby,nonmarkov,polonica} As already
mentioned, the coefficient $\left( 1-B_{1}\frac{a}{\xi }\right) $
introduces
finite-size effects through the ratio $a/\xi $ and the parameter
$B_{1}$
takes its value depending on the nature of the boundary.\cite%
{brenner,saarloos,libchaber}

\section{Confined diffusion of a finite sized particle}

Once introduced the general aspects of the model, we will analyze the
particular case of a particle moving in a viscoelastic medium in which
the
effects of constrainng are also present. Finally, we will compare our
results with experiments.

Substituting the harmonic force (\ref{FB1}), the relation for the
effective
diffusion coefficient (\ref{Stokes-Einstein}) and the effective
mobility (%
\ref{friccion-pared}), the generalized difusion equation
(\ref{Smol-Osc})
takes the form
\begin{equation}
\frac{\partial \rho }{\partial t}=\beta ^{-1}(t)\left[
\frac{k_{B}T}{m}%
\nabla ^{2}\rho +\epsilon \omega _{0}^{2}\nabla \cdot (\vec{x}\rho
)\right] .
\label{Smol-Osc2}
\end{equation}

A non-stationary solution of Eq. (\ref{Smol-Osc2}) can be obtained by
first
introducing the dimensionless coordinates
$\tilde{\vec{x}}=a^{-1}\vec{x}$
and the dimensionless time $\tilde{t}=t_{0}^{-1}t$, with $t_{0}=\beta
_{0}^{-1}$, and scaling time with
\begin{equation}
\tau =\int_{0}^{\tilde{t}}\tilde{\beta}^{-1}(\tilde{t}^{\prime
})\,d\tilde{t}%
^{\prime }.  \label{tau}
\end{equation}%
Assuming a probability distribution of the form: $\rho
(\tilde{\vec{x}},\tau
)=f(\tau )exp\left[ -\tilde{\vec{x}}^{2}/A(\tau )\right] $, one arrives
at
the solution
\begin{eqnarray}
\rho &=&\frac{\rho _{0}}{\left( 1-e^{-{2\omega
_{0}^{2}\tau }/{\alpha \omega _{T}^{2}}}\right) ^{\frac{3}{2}}}\cdot
\,\,\,\,\,\,\,\,\,\,\,\,\,\,\,\,\,\,\,\,\,\,\,\,\,\,\,\,\,\,\,\,\,\,\,\,\,\,%
\,\,  \label{Rhocompleta} \\
&&exp\left[ {-\frac{\omega _{0}^{2}\beta _{0}^{2}}{\omega _{T}^{4}}%
\frac{{\tilde{\vec{x}}^{2}}}{2\alpha ^{2}}\left( 1-e^{-\frac{2{\omega
_{0}^{2}}}{\alpha {\omega _{T}^{2}}}\tau }\right) ^{-1}}\right] ,
\nonumber
\end{eqnarray}%
in which $\rho _{0}$ is a normalization constant and we have defined
the
frequency $\omega _{T}^{2}=k_{B}T/ma^{2}$ and, according to Eq. (\ref%
{friccion-pared}), $\alpha $ is given by
\begin{equation}
\alpha =\left( 1-B_{1}\frac{a}{\xi }\right) .  \label{exponentealfa}
\end{equation}%
Eq. (\ref{Rhocompleta}) gives the probability distribution for a single
particle in the case of constrained diffusional
motion.\cite{cromosomas} It
leads to the following expression for the mean square displacement of
the
particle
\begin{equation}
\langle \tilde{\vec{x}}^{2}\rangle (\tau )= \frac{3\omega _{T}^{4}}{%
\beta _{0}^{2}\omega _{0}^{2}}\alpha ^{2}\left[ 1+{\text{coth}}\left(
\frac{%
2\omega _{0}^{2}}{\alpha \omega _{T}^{2}}\tau \right) \right] ^{-1},
\label{xx-cromosomas}
\end{equation}%
which can be obtained by using the definition $\langle \tilde{\vec{x}}%
^{2}\rangle =\int {\tilde{\vec{x}}^{2}\rho \,d\tilde{\vec{x}}}$. In
Fig.
\textbf{2}, we show a fitt (with Eq. (\ref{xx-cromosomas})) of the
experimental data of constrained diffusion of chromatin inside the
nucleus
of living cells, see caption of the figure for details.

\begin{figure}[tbp]
{}
\par
\centering \mbox{\resizebox*{7.5cm}{!}{\includegraphics{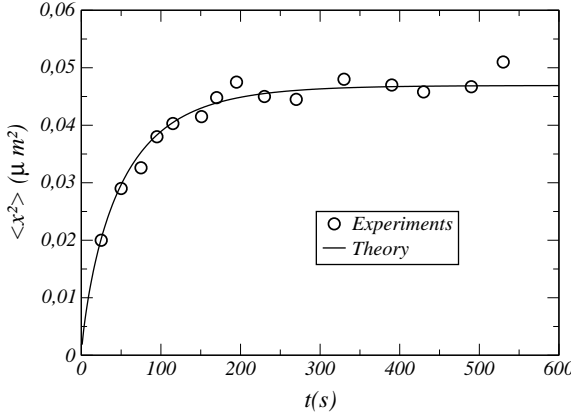}} }
\par
{\footnotesize {\ } \vspace{0.0cm} }
\caption{ Fitting of experimental data (circles), corresponding to
constrained diffusion in living cells given in Fig. \textbf{2b} of Ref.
[12], by means of the MSD obtained in Eq. (\protect\ref%
{xx-cromosomas}) with ${\omega^{2}_T}/{\omega^{2}_0} = 4/5$, $B_2=1/6$
and $6D_0\beta^{-1}_0\alpha
t_0^{-\alpha{\omega^{2}_T}/{\omega^{2}_0}}\sim 0.02 \mu m^2$. The values of the parameters are given in the text.}
\label{msd}
\end{figure}

Experiments measuring the mean square displacement of Brownian
particles in
\textit{in vitro} F-actin networks shown a power law behavior
characteristic
of subdiffusion in which the exponent is a function of the
particle-to-network-size ratio $a/\xi $ (see
Fig. \textbf{1}).\cite{wong} To explain the origin
of this short-time power law behavior of the MSD, we will first obtain
the
explicit dependence of $\tau (\tilde{t})$ at short times. To achieve
this objective, it is convenient to analyze the
evolution equation for the relaxation function $\chi (\tilde{t})=\int
{\vec{x%
}\cdot\vec{x}_{0}\rho d\vec{x}}$:\cite{kubo,nonmarkov}
\begin{equation}
\frac{d}{d\tilde{t}}\chi (\tilde{t})=-\frac{\omega _{0}^{2}}{\alpha
\omega
_{T}^{2}}\tilde{\beta}^{-1}(\tilde{t})\chi (\tilde{t}).
\label{evolRelajacion}
\end{equation}%
Eq. (\ref{evolRelajacion}) can be derived by using Eqs.
(\ref{friccion-pared}%
) and (\ref{Smol-Osc2}).\cite{adelman,nonmarkov} Then, by taking
into account Eqs. (\ref{tau}) and (\ref{evolRelajacion}) one may obtain
the
relation
\begin{equation}
\tau =\ln \left[ \frac{\chi (\tilde{t}_{0})}{\chi (\tilde{t})}\right]
^{\alpha \omega _{T}^{2}/\omega _{0}^{2}},  \label{tau1}
\end{equation}%
which can be expressed in a more convenient way by expanding $\chi
(\tilde{t}%
)$ up to first order in $\tilde{t}$: $\chi (\tilde{t})\simeq \chi
(\tilde{t}%
_{0})(1-B_{2}\tilde{t})+O(\tilde{t}^{2})$, with $B_{2}$ a parameter
characterizing the type of relaxation. The resulting expression is
$\tau
\simeq \ln \left[ 1+B_{2}\tilde{t}\right] ^{\alpha \omega
_{T}^{2}/\omega
_{0}^{2}}$. After expanding the logarithm around the unity, one finally
finds
\begin{equation}
\tau \sim B_{2}\,\tilde{t}^{\alpha \omega _{T}^{2}/\omega _{0}^{2}}.
\label{tau2}
\end{equation}%
With this expression, the relaxation function is found to be
\begin{equation}
\chi (\tilde{t})\sim \chi (\tilde{t}_{0})e^{-\frac{\omega
_{0}^{2}}{\alpha
\omega _{T}^{2}}B_{2}\tilde{t}^{\alpha \omega _{T}^{2}/\omega
_{0}^{2}}},
\end{equation}%
which is a stretched exponential characteristic of the relaxation of
systems with strong spatial inhomogeneities.

The MSD can now be calculated by using the short time approximation of
Eq. (%
\ref{Rhocompleta}), which can be obtained by expanding the exponential
$e^{-%
\frac{2\omega _{0}^{2}}{\alpha \omega _{T}^{2}}\tau }$ up to first
order in $%
\tau $, from where one obtains%
\begin{equation}
\rho (\tilde{\vec{x}},\tau )\cong \frac{\rho _{0}}{\left( {2\omega
_{0}^{2}\omega _{T}^{-2}\alpha }^{-1}{\tau }\right)
^{\frac{3}{2}}}\,\,e^{%
\left[ {-\frac{\beta _{0}^{2}}{4\alpha \omega
_{T}^{2}}\frac{{\tilde{\vec{x}}%
^{2}}}{\tau }}\right] }.  \label{Rho-short}
\end{equation}
Then, from Eqs. (\ref{Rho-short}) and (\ref{tau2}) one obtains in the
original variables
\begin{equation}
\langle \vec{x}^{2}\rangle (t)\simeq 6D_{0}\beta _{0}^{-1}\alpha
B_{2}\left[
\frac{t}{t_{0}}\right] ^{\alpha \omega _{T}^{2}/\omega _{0}^{2}},
\label{xx-3}
\end{equation}%
where $D_{0}=k_{B}T/m\beta _{0}$ is the usual expression for the
diffusion
coefficient in terms of the Stokes friction coefficient $\beta _{0}$.
The
result clearly shows that the finite size of the particles affect the
expression of the MSD through the factor $a/\xi $.

In Fig.\textbf{1}, we compare our results for the MSD with the
experimental
data reported in Ref. [8]. The slopes of the dotted lines were
obtained by using Eq. (\ref{xx-3}) and the relation $\alpha
=1-B_{1}a/\xi $
with $B_{1}=2/3$ (a value similar to that obtained from
hydrodynamics\cite%
{brenner,dufresne}) $D_{0}\sim 10^{12}cm^{2}s^{-1}$, $\beta _{0}\sim
10^{6}s^{-1}$, $\omega _{T}\sim 2.25\cdot 10^{3}s^{-1}$ and $\omega
_{0}\sim
2\cdot 10^{3}s^{-1}$ for the different values of $\alpha /\xi $
indicated in
the caption. With these values of the constants, the experimental
magnitude
of the MSD were well fitted by assuming that $\alpha \simeq \lbrack
1-(2/3)(a/\xi )]\exp [-5(a/\xi )]$. Finally, in figure \textbf{1}
solid lines represent the fit of the data using Eq.
(\ref{xx-cromosomas}) and thus
indicating the effects of confinement due to the material surrounding
the particle.

At short times, the viscoelastic properties of the intracellular medium
become manifest through the effective diffusion coefficient $D(t)$
given in
Eq. (\ref{Stokes-Einstein}), and which can explicitly be obtained by
taking
the time derivative of Eq.
(\ref{xx-3}).\cite{libchaber,bickel,dufresne}
After using the definition of $\langle \vec{x}^{2}\rangle $ and Eqs.
(\ref%
{friccion-pared}), (\ref{Smol-Osc2}) and (\ref{exponentealfa}), we have
identified $B_{2}=\alpha ^{-1}{\omega _{0}^{2}}/{\omega _{T}^{2}}$ and
the
explicit expression for the dimensionless function
$\tilde{\beta}^{-1}(t)$
which incorporates memory and finite-size effects
\begin{equation}
\tilde{\beta}^{-1}(t)=\left[ \frac{t}{t_{0}}\right] ^{{\alpha \omega
_{T}^{2}%
}/{\omega _{0}^{2}}-1}.  \label{Zetaeff}
\end{equation}

The relation between the effective mobility per unit mass (\ref%
{friccion-pared}) with the effective viscosity of the intracellular
medium
can be established by realizing that $\beta (t)$ represents a temporal
average of the form\cite{balescu,zwanzig,kubo}
\begin{equation}
\beta (t)=\int_{\tau _{D}}^{t}{\zeta (z)dz},  \label{Zetaeff-Av}
\end{equation}%
where $\zeta (t)$ is a memory function related with the
frequency-dependent
effective viscosity through the inverse Laplace transform of $\zeta
(\omega
)\sim \eta (\omega )$,\cite{weitz82,mason2000,SteadyState} and we have
introduced $\tau _{D}$ as a cut-off time. After calculating $\zeta (t)$
from
Eqs. (\ref{Zetaeff}) and (\ref{Zetaeff-Av}), the Laplace transform of
the
result leads to the scaling law for the effective viscosity
\begin{equation}
\eta (\omega )\sim \omega ^{{\alpha \omega _{T}^{2}}/{\omega
_{0}^{2}}-1}.
\label{Etaeff}
\end{equation}%
From this expression, one may calculate the dependence of the shear
modulus $%
G^{\prime \prime }(\omega )=\omega \eta (\omega )$, as
\begin{equation}
G^{\prime \prime }\sim \omega ^{{\alpha \omega _{T}^{2}}/{\omega
_{0}^{2}}}.
\label{G}
\end{equation}%
As in the experiments,\cite{wong,gittes} this expression shows that the
shear modulus $G^{\prime \prime }(\omega )$ inferred through the
generalized
Stokes-Einstein relation $D(\omega )=\frac{k_{B}T}{6\pi a\eta (\omega
)}$,
depends in general on the ratio $a/\xi $.\cite{wong,mason2000}

The results here obtained can also be used to explain the subdiffusive
behavior, $\langle x^{2}\rangle _{exp}\sim t^{3/4}$, found in
experiments
with passively diffusing particles in living cells.\cite{caspi} The
form of
the MSD with ${\alpha \omega _{T}^{2}}/{\omega _{0}^{2}}=3/4$, and Eqs.
(\ref%
{Zetaeff}), (\ref{Etaeff}) and (\ref{G}) then imply the behaviors
$\beta
(t)\sim t^{-1/4}$, $\eta (\omega )\sim \omega ^{-1/4}$ and $G^{\prime
\prime
}\sim \omega ^{3/4}$, in good agreement with the results of Refs. [11].
The experiments were performed with particles having diameters in
the range $a\simeq 2-3\mu m$.\cite{caspi} Thus, our relation ${a}/{\xi
}%
\sim 2/5$ implies that the characteristic length of the cytoskeleton
network
is $\xi \simeq 5-7.5\mu m$ which lies within the same range of values
as
the actin persistence length $\xi _{actin}\simeq 5-20\mu
m$.\cite{caspi}

\section{Discussion}

We have shown that the observed finite-size effects in the
anomalous transport of particles in the intracellular environment can
be
explained on the grounds of a diffusion model which incorporates the
viscoelastic nature of the intracellular medium, and the characteristic
lengths associated to the size of the particles and of the polymer
network
constituting the cytoskeleton.

The proposed diffusion equation incorporates memory effects through
time dependent coefficients which enables one to calculate the mean-square
displacement from which the role played by the mentioned finite-effects
effects can be analyzed. These effects correspond to the size of the
particles and the characteristic length of the polymer
network and of the size of the system, i.e., our description may
include the effects of confinement. As a consequence, by assuming that the
constants take values typical of experiments, we have found good
agreement between our theoretical results and the experimental data taken
from microrheological experiments with viscoelastic fluids similar to the
intracellular
medium. Good agreement has also been found in describing confined
diffusion of
chromatin within the nucleus of live cells.

In summary, the scheme
presented can be applied to explain the observed subdiffusive
behavior of particles passively diffusing in the cell.

We thank R. Rodriguez and D. Reguera for valuable discussions. This
work was partially supported by the grant DGAPA-UNAM IN108006 (ISH) and
DGICYT of the Spanish Government
under Grant No. PB2002-01267.

\end{document}